\documentclass{moriond}

\usepackage{amsmath}
\usepackage{amssymb}
\newcommand{\Photo}{\includegraphics[height=35mm]{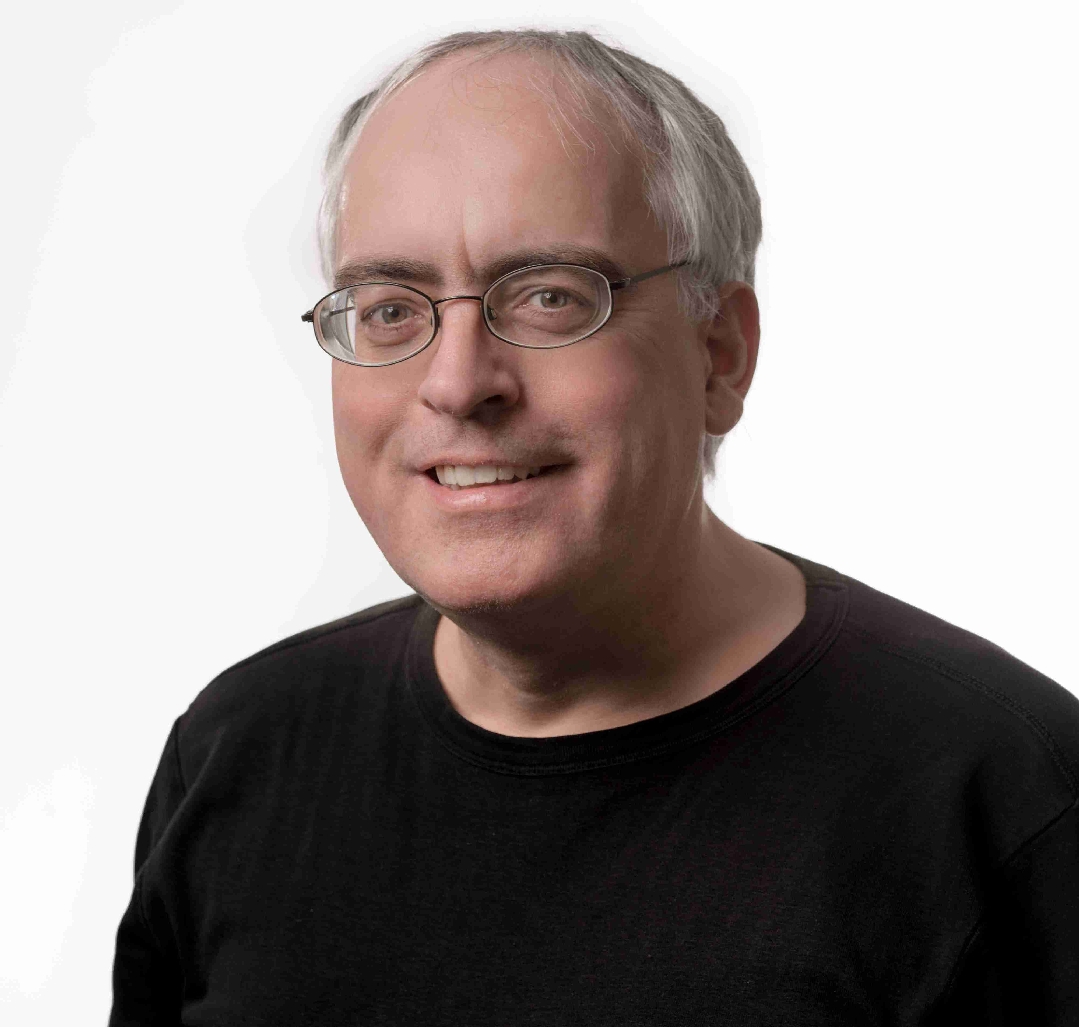}}

\newcommand{\kk}{$K\!-\ov{\!K}{}\,$}
\newcommand{\kkm}{\kk mixing}

\newcommand{\beqin}[1]{$ #1 $}
\newcommand{\bra}[1]{\langle #1|} 
\newcommand{\ket}[1]{|#1\rangle}
\newcommand{\braket}[2]{\langle #1|#2\rangle}

\newcommand{\lt}{\left}
\newcommand{\rt}{\right}

\newcommand{\tev}{\,\mbox{TeV}}
\newcommand{\gev}{\,\mbox{GeV}}

\newcommand{\Kbar}{\,\overline{\!K}}

\newcommand{\nn}{\nonumber \\}
\newcommand{\no}{\nonumber}
\newcommand{\ov}{\overline}

\newcommand{\imag}{{\rm Im}\,}
\newcommand{\real}{{\rm Re}\,}

\newcommand{\fig}[1]{Fig.~\ref{#1}}
\newcommand{\eq}[1]{Eq.~(\ref{#1})}

\begin{document}
\vspace*{4cm}
\title{$\mathbf{\epsilon_K^\prime/\epsilon_K}$: 
         Standard Model and Supersymmetry}

\author{Ulrich Nierste}

\address{Institute for Theoretical Particle Physics (TTP)\\
        Karlsruhe Institute of Technology (KIT)         \\
        76131 Karlsruhe, Germany\\
        E-mail: ulrich.nierste@kit.edu}

\maketitle\abstracts{I give a pedagogical introduction into
  flavour-changing neutral current interactions of kaons and their role
  to reveal or constrain physics beyond the Standard Model (SM).  Then I
  discuss the measure $\epsilon_K^\prime$ of direct CP violation in
  $K\to \pi\pi$ decays, which deviates from the SM prediction by
  $2.8\sigma$. A supersymmetric scenario with flavour mixing among
  left-handed squarks can accomodate the measured value of
  $\epsilon_K^\prime$ even for very heavy sparticles, outside the reach of
  the LHC. The considered scenario employs mass splittings among the
  right-handed up and down squarks (to enhance $\epsilon_K^\prime$) and
  a gluino which is heavier than the left-handed strange-down mixed
  squarks by at least a factor of 1.5 (to suppress excessive
  contribution to $\epsilon_K$, the measure of indirect CP violation).
  The branching ratios of the rare decays $K^+ \to \pi^+ \nu \bar\nu$ and
  $K_L \to \pi^0 \nu \bar\nu$, to be measured by the NA62 and KOTO-step2
experiments, respectively, are only moderately affected. These
measurements have the capability to either falsify the model or to 
constrain the CP phase associated with strange-down squark mixing accurately.  
}

\section{Basics}
The gauge interactions of the Standard Model (SM) are \emph{flavour
  blind}, meaning that they treat all three fermion generations
equally. The Yukawa interaction between fermions and Higgs field, however,
involves different coupling strengths of the Higgs field to fermions of
different generations, encoded in the complex $3\times 3$ Yukawa
matrices. The products of these matrices and the vacuum expectation value
of the Higgs field constitute the fermion mass matrices.
The diagonalisation of the mass matrices involves unitary rotation of the
fermion fields which cancel everywhere except in the couplings of
the $W$ boson to fermions. The unitary matrix appearing in the
charged-current $W$ couplings to quarks of different generations is the
Cabibbo-Kobayashi-Maskawa (CKM) matrix $V$.

\begin{figure}[t]
\centering
\includegraphics[height=2.5cm]{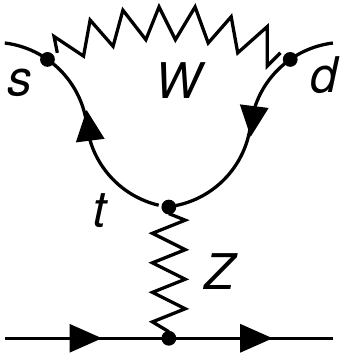}
\hspace{15mm}
\includegraphics[height=3cm]{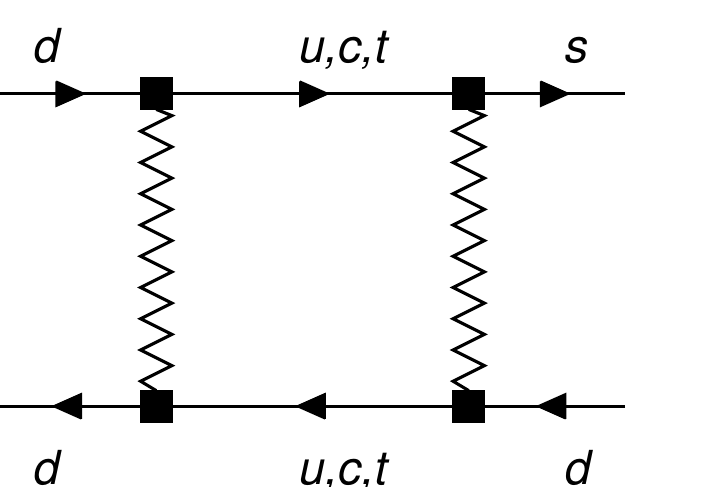}
\hspace{15mm}
\includegraphics[height=3cm]{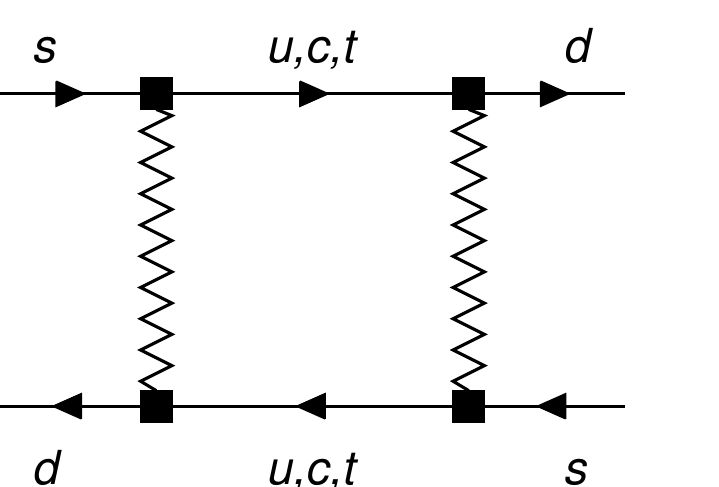}
\caption{Left and middle: sample diagrams of electroweak $\Delta S=1$
  penguins and boxes which contribute to the low-scale value of the
  Wilson coefficient of $Q_8$ (through renormalisation group evolution from
  other electroweak coefficients). Right: $\Delta S=2$ box diagram of
  \kkm. \label{fig:diagr} }
\end{figure}

$W$ couplings necessarily change the fermion electric charge by one
unit.  Thus flavour-changing neutral current (FCNC) transitions are
forbidden at tree-level in the SM and only proceed through loops.
Decays of charged ($K^+\sim \bar s u$) and neutral ($K^0\sim \bar s d$)
kaons can be used to study the $s\to d$ FCNC transitions. Here the FCNC
transition of interest is either the decay amplitude ($|\Delta S|=1$
transition where $S$ means \emph{strangeness}) or the \kkm\ amplitude, a
$|\Delta S|=2$ process (see \fig{fig:diagr} for sample diagrams).
\kkm\ implies that $K^0\sim \bar s d$ and $\Kbar{}^0 \sim s\bar d$ are 
not the physical eigenstates of neutral kaons. Instead,
these are $K_L$ and $K_S$, the long-lived and short-lived neutral kaons,
which are linear combinations of the flavour eigenstates $K^0$
and $\Kbar{}^0$. 

Within the SM, FCNC transitions of kaons are not only loop suppressed:
The CKM-favoured $s\to d$ transitions proportional to $V_{us}V_{ud}^*$
and $V_{cs}V_{cd}^*$ involve an up and a charm quark on the internal
line, respectively. These two contributions almost cancel each other
perfectly, owing to $V_{us}V_{ud}^*\approx -V_{cs}V_{cd}^*$ and the
smallness of $m_c^2-m_u^2$ compared to $M_W^2$, rendering the up and
charm loop integral almost equal in size. This feature, the
\emph{Glashow-Iliopoulos-Maiani mechanism}, makes kaon FCNC processes
sensitive to virtual effects of the heavy top quark and of potential new
particles predicted by theories beyond the SM. To this end charge-parity
(CP) violating observables and branching ratios of rare decays are of
key importance. 

If CP were a good symmetry, the neutral kaon mass eigenstates
$\ket{K_L}$ and $\ket{K_S}$ should coincide with the CP eigenstates
$(\ket{K^0}\pm\ket{\Kbar{}^0})/\sqrt2$. The two-pion states
$\ket{\pi^+\pi^-}$ and $\ket{\pi^0\pi^0}$ are CP even and indeed $K_S$
mesons predominantly decay into two pions, while $K_L$ prefers
three-pion decays, suggesting that $K_S$ and $K_L$ are CP-even and
CP-odd, respectively.  But the 1964 discovery of $K_L\to \pi\pi$ decays
has shown that $K_L$ is not a CP eigenstate and therefore established CP
violation \cite{Christenson:1964fg}. Today we know that the discovered
effect is related to the top quark in the box diagram in
\fig{fig:diagr}. That is, the decay of particle with a mass of
0.5\gev\ revealed the virtual effect of a particle which is roughly 350
times heavier, which testifies to the tremendous discovery potential of
kaon FCNCs!
  
The strong interaction poses the main challenge for theoretical predictions
of kaon decays. Short-distance QCD effects can be calculated in
perturbation theory, from e.g.\ Feynman diagrams with gluons dressing 
the diagrams of \fig{fig:diagr}. To separate these from the
non-perturbative long-distance QCD effects, one sets up an effective
hamiltonian which reads  
\begin{eqnarray}
 \mathcal{H}_{\textrm{eff}}^{ | \Delta S|=1} &= 
 \frac{G_F}{\sqrt{2}} \lambda_u \sum_{i=1}^{10} Q_i (\mu) \left( z_i
 (\mu) + \tau y_i (\mu) \right) + \textrm{H.c.}
\label{eq:hamilton}
\end{eqnarray} 
for the case of $K\to \pi\pi$ decays. Here $\lambda_u = V_{us}^{\ast}
V_{ud} $ and $\tau = - V_{ts}^{\ast} V_{td} / \left( V_{us}^{\ast}
V_{ud} \right)$ and $G_F$ is the Fermi constant.  Pictorially, the
operators $Q_i$ are found by contracting the $W$ lines and loops in the
Feynman diagrams to a point; they describe effective four-quark
interactions.  The operator basis {comprises} ten operators which are
defined in Ref.~\cite{Buras:1993dy}. In the following we will discuss 
\begin{equation}
Q_6=\ov s^j \gamma_\mu (1-\gamma_5) d^k 
                \sum_q \ov{q}{}^k\gamma^\mu (1+\gamma_5)q^j \quad 
\mbox{and}\quad
Q_8=\frac32\ov s^j \gamma_\mu (1-\gamma_5)d^k\sum_q 
        e_q \ov{q}{}^k\gamma^\mu (1+\gamma_5)q^j \label{eq:ops}
\end{equation}
with colour indices $j,k$ and $e_q$ being the charge of quark $q$.  The
Wilson coefficients $z_i$ and $y_i$ in \eq{eq:hamilton} are calculated
from the Feynman diagrams and comprise the short-distance physics, in
particular they depend on the masses of the heavy $W$, top and potential
new-physics particles. The coefficients depend on the renormalisation
scale, the diagrams in \fig{fig:diagr} determine the values of the
coefficients at the high scale of order of the $W$ or top mass. Their
values at the low scale around 1\gev\ relevant for kaon physics are
found by solving the renormalisation group equations.  \fig{fig:diagr}
shows diagrams contributing to $y_{7,9}$ at the high scale. As a result
of the renormalisation group evolution, the low-scale values of the
$y_i$ are linear combinations of the high-scale values, which renders
the coeffient of interest, $y_8$, non-zero and important for
$\epsilon_K^\prime$. The Wilson coefficients are known to
next-to-leading order in the strong 
coupling constant $\alpha_s$ \cite{nlo}. 

The effective $\Delta S=2$ hamiltonian needed to describe \kkm\ is much
simpler and contains only one operator $\bar s\gamma_\mu(1-\gamma_5)
d\, \bar s\gamma^\mu(1-\gamma_5) d$.
 
In order to predict $K\to \pi\pi$ decay amplitudes one must calculate
the hadronic matrix elements $\braket{\pi\pi}{Q_i|K}$
with non-perturbative methods such as lattice gauge theory. To this end
it is advantageous to switch to eigenstates of the strong isospin $I$:
\begin{eqnarray}
\ket{\pi^0 \pi^0 } =  
      \sqrt{\frac{1}{3}} \, \ket{\lt(\pi \pi \rt)_{I=0}}  
    - \sqrt{\frac{2}{3}} \, \ket{\lt(\pi \pi \rt)_{I=2}}, &&\quad
\ket{\pi^+ \pi^- } =  
      \sqrt{\frac{2}{3}} \, \ket{\lt(\pi \pi \rt)_{I=0}}  
    + \sqrt{\frac{1}{3}} \, \ket{\lt(\pi \pi \rt)_{I=2}}  . \no
\end{eqnarray}
(Bose symmetry forbids the $I=1$ state.) Strong isospin is an excellent
symmetry of QCD and was exact, if up and down quark had the same mass.

In this talk I present new analyses of $\epsilon_K^\prime$ in the SM
and its minimal supersymmetric extension (MSSM) from the papers 
\cite{Kitahara:2016nld,Kitahara:2016otd,Crivellin:2017gks}. This
research was stimulated by a breakthrough of the RBC and UKQCD
collaborations in the calculation of the matrix elements 
$\braket{\lt(\pi \pi \rt)_{I=0}}{Q_i|K^0}$ with lattice gauge 
theory \cite{Bai:2015nea}. The results of Ref.\ \cite{Crivellin:2017gks}
are further covered by Teppei Kitahara's talk at this
conference \cite{Kitahara:2017uva}.  

\section{Indirect and direct CP violation}
We express the CP-violating quantities of interest in terms of
the decay amplitudes $A(K_{L,S}\to (\pi \pi)_{I=0,2})$. 
Indirect CP violation 
(stemming from the $\Delta S=2$ box diagrams) is quantified by 
\begin{equation}
\epsilon_K\equiv \frac{A (K_L\to (\pi \pi)_{I=0})}{
                            A (K_S\to (\pi \pi)_{I=0})} =
   (2.228\pm 0.011)\cdot 10^{-3}\cdot e^{i(0.97\pm0.02)\pi/4}.
\end{equation}
The measure of direct CP violation, which originates from the $\Delta
S=1$ kaon decay amplitude and is calculated from
$\mathcal{H}_{\textrm{eff}}^{ | \Delta S|=1}$ in \eq{eq:hamilton}
is\footnote{Accidentally, $\epsilon_K^\prime/\epsilon_K$ is essentially
  real.}
\begin{equation}
\epsilon_K^\prime \simeq
\frac{\epsilon_K}{\sqrt{2}} \lt[ 
    \frac{  A(K_L\to (\pi\pi)_{I=2}) }{ A(K_L\to (\pi\pi)_{I=0})}
    - 
    \frac{  A(K_S\to (\pi\pi)_{I=2}) }{ A(K_S\to (\pi\pi)_{I=0})}
    \rt]
= (16.6\pm2.3)\cdot 10^{-4} \cdot \epsilon_K .\label{eq:exp}
\end{equation}
This experimental result was established in 1999 and constituted the
first measurement of direct CP violation in any decay \cite{epspexp}.
In kaon physics one usually adopts the standard phase convention of the
CKM matrix, so that the dominant tree-level $\bar s\to \bar d u \bar u $
amplitude proportional to $V_{ud} V_{us}^*$ is real. CP violation
requires the interference of this amplitude with another amplitude
having a different weak phase, which in the SM stems from another
combination of CKM elements. Diagrams with internal top quark like the
penguin diagram shown in \fig{fig:diagr} are proportional to $V_{td}
V_{ts}^*$ and charm loops contribute to both CKM structures, owing to
CKM unitarity, $V_{cd} V_{cs}^*=-V_{td} V_{ts}^*-V_{ud} V_{us}^*$. Thus
SM predictions for CP-violating quantities in $K\to \pi\pi$  decays 
all involve 
\begin{equation}
  \tau = - \frac{V_{td}V_{ts}^*}{V_{ud}V_{us}^*} \sim (1.5 - i 0.6)\cdot
  10^{-3}. \label{eq:deftau}
\end{equation}
But   $\epsilon_K^\prime$ is not only suppressed by the smallness of
$\imag\tau$, strong interaction effects also render the amplitudes in
the numerators in \eq{eq:exp} much smaller than the
denominators. Experimentally one finds for $A_I\equiv A(K^0\to (\pi
  \pi)_{I})$:\footnote{By convention the CP-conserving, strong phase $\delta_I$ is
    factored out, so that
    $\braket{(\pi\pi)_I}{\mathcal{H}_{\textrm{eff}}^{ | \Delta S|=1}
    | K^0}\equiv A_I \exp(i\delta_I)$. The phase $\delta_I$ is generated
by final-state rescattering of the pions. A non-zero $\imag A_I$
therefore only arises from the non-zero CP phase $\arg\tau$.}   
\begin{equation}
\real A_0 =  \left( 3.3201\pm 0.0018 \right)\times
10^{-7}~\gev,\qquad 
\real A_2 =  \left(1.4787 \pm 0.0031\right) \times 10^{-8}~\gev,
\label{eq:a02}
\end{equation}
with a ratio $\real A_0/\real A_2\approx 22$! This feature is called 
\emph{$\Delta I=1/2$ rule}, because $I$ changes by half a unit in 
$K_{L,S}\to (\pi\pi)_{I=0}$. 

The master equation for \beqin{\epsilon_K^\prime/\epsilon_K} (see e.g.\
Ref.~\cite{Buras:2015yba})
reads:
\begin{eqnarray}
  \frac{\epsilon_K^\prime}{\epsilon_K} = \frac{\omega_{+}}{\sqrt{2}
    {|}\epsilon_K^{\textrm{exp}}{|} \real A_0^{\textrm{exp}} }
  \left\{ \frac{\imag A_2 }{\omega_{+}} - \left( 1-
      \hat{\Omega}_{\textrm{eff}} \right) \imag A_0 \right\} .
\label{eq:mas}
\end{eqnarray}
Here \beqin{\omega_{+}\simeq \frac{\real A_2}{\real A_0} = (4.53 \pm
  0.02)\cdot10^{-2}} is determined from the charged counterparts of
$\real A_{0,2}$ and \beqin{\hat{\Omega}_{\textrm{eff}} = (14.8\pm
  8.0)\cdot 10^{-2}} quantifies isospin breaking. One also takes
$|\epsilon_K^{\textrm{exp}}|$ and $\real A_0^{\textrm{exp}} $ from
experiment, as indicated. The theoretical challenge is the calculation
of $\imag A_{0,2}$ with non-perturbative methods.  Within the SM
\beqin{\imag A_0} is dominated by gluon penguins, with
roughly 2/3 stemming from the matrix element \beqin{\bra{(\pi\pi)_{I=0}}
  Q_6 \ket{K^0}} (with the operator $Q_6$ of \eq{eq:ops}), while about 3/4 of
the contribution to \beqin{\imag A_2} stems from
\beqin{\bra{(\pi\pi)_{I=2}} Q_8 \ket{K^0}}.  Lattice-gauge theory has
\beqin{\bra{(\pi\pi)_{I=2}} Q_8 \ket{K^0}} (and thereby $\imag A_2$)
under good control for some time \cite{Blum:2011ng}, while reliable
lattice calculations of \beqin{\bra{(\pi\pi)_{I=0}} Q_6 \ket{K^0}} have
become possible only recently \cite{Bai:2015nea}.  Using these matrix
elements from lattice QCD we find \cite{Kitahara:2016nld}
\begin{equation}
  \frac{\epsilon_{K}^\prime}{\epsilon_{K}}  =
  \left( {1.06}\pm  {4.66_{\rm{Lattice}}} \pm {1.91_{\rm{NNLO}}} \pm  
    {0.59_{\rm{IV}}} \pm  { 0.23_{m_t}} \right) \times 10^{-4},
\label{eq:smnlo}
\end{equation}
a value which is $2.8\,\sigma$ below the experimental result in
\eq{eq:exp}. The various sources of errors are indicated by the
subscripts: The largest uncertainty stems from the hadronic matrix
elements calculated with lattice QCD. The next error is the perturbative
uncertainty from the unknown next-to-next-to-leading (NNLO) QCD
corrections. ``IV'' denotes strong-isospin violation (stemming e.g.\
from $m_u\neq m_d$) and the last error comes from the error in $m_t$.

This result, obtained with a novel compact solution of the
renormalization group equations, agrees with the one in
Ref.~\cite{Buras:2015yba}.  The quoted lattice results are consistent
with earlier analytic calculations in the large-$N_c$ ``dual QCD''
approach \cite{bbg}. Thus lattice gauge theory is currently starting to
resolve a long-standing controversy about $\imag A_0$ between the
large-$N_c$ \cite{bbg} and chiral perturbation theory \cite{chpt}
communities. While the latter method can reproduce the large-$N_c$
values, it can likewise easily accomodate the experimental range in
\eq{eq:exp}.

\section{$\mathbf{\epsilon_K^\prime}$ in the MSSM}
The large factor $1/\omega_{+}$ multiplying $\imag A_2$ in \eq{eq:mas}
renders $\epsilon_K^\prime/\epsilon_K$ especially sensitive to new
physics in the $\Delta I=3/2 $ decay $K\to (\pi\pi)_{I=2}$. This feature
makes $\epsilon_K^\prime/\epsilon_K$ special among all FCNC processes.
However, it is difficult to place a large effect into
$\epsilon_K^\prime$ without overshooting $\epsilon_K$: The SM
contributions to both quantities depend on the CKM combination 
$\tau$ in \eq{eq:deftau} as
\begin{equation}
\epsilon_K^{\prime\,\rm SM} \propto \imag \tau
  \qquad \mbox{and}\qquad
  \epsilon_K^{\rm NP} \propto \imag \tau^2 .
\label{eq:scsm}
\end{equation}
In new-physics scenarios $\tau $ is replaced by some new $\Delta S=1$
parameter $\delta$ 
and the new-physics contributions scale as 
\begin{equation}
\epsilon_K^{\prime\, \mathrm{NP}} \propto \imag \delta\qquad 
\mbox{and}
\qquad
\epsilon_K^{\mathrm{NP}} \propto \imag \delta^2. \label{eq:scnp}
\end{equation}
If new-physics enters through a loop with super-heavy particles, the
only chance to have a detectable effect in $\epsilon_K^\prime$ is a
scenario with $|\delta|\gg |\tau|$. Thus if $\epsilon_K^{\prime\,\rm
  NP}\sim \epsilon_K^{\prime\,\rm SM}$ one expects $\epsilon_K^{\rm
  NP}\gg \epsilon_K^{\rm SM}$, in contradiction with the experimental
value.  Thus large effects in $\epsilon_K^\prime$ from loop-induced new
physics are seemingly forbidden. Many studies of $\epsilon_K^\prime$
indeed involve new-physics scenarios with tree-level contributions to
$\epsilon_K^\prime$ \cite{nptree}, in which the requirement $|\delta|\gg
|\tau|$ can be relaxed.

\begin{figure}[t]
\centering
\includegraphics[height=2cm]{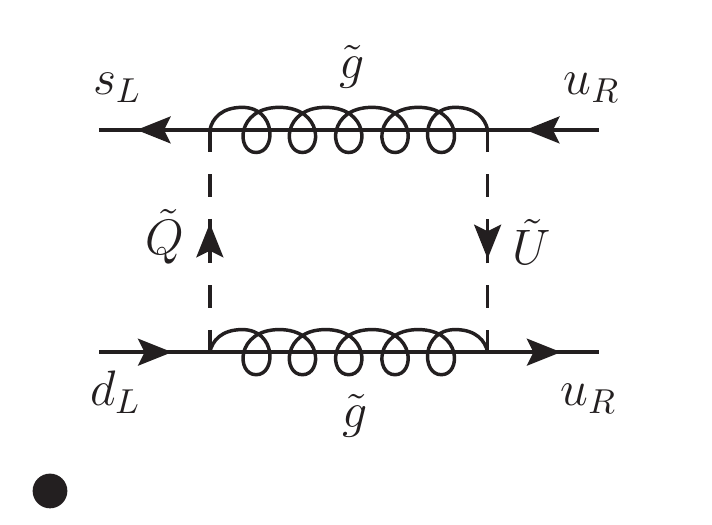}\hspace{5mm}
\includegraphics[height=2cm]{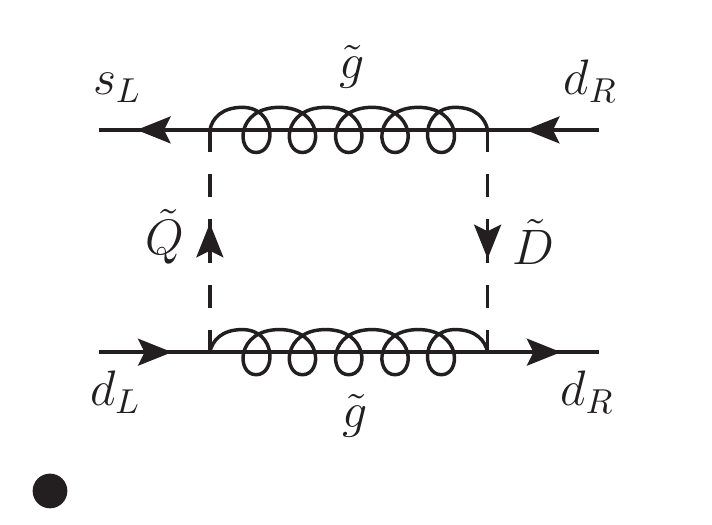}\hspace{5mm}
\includegraphics[height=2cm]{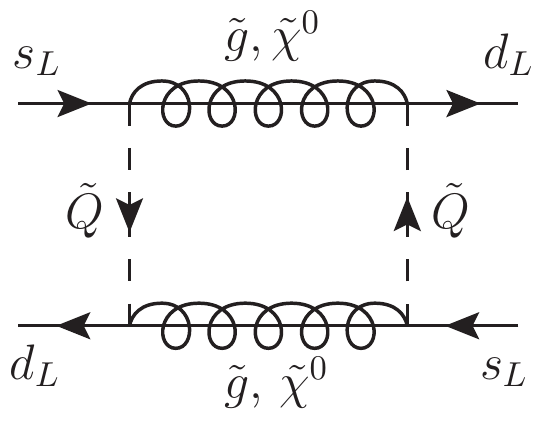}\hspace{5mm}
\includegraphics[height=2cm]{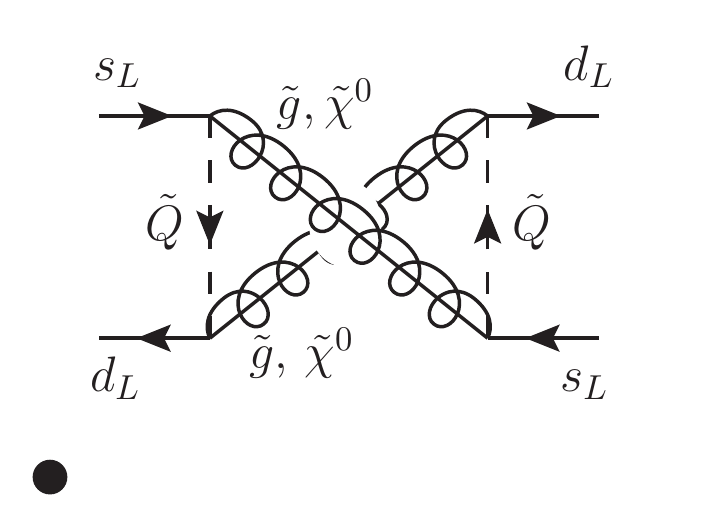}\hspace{5mm}
\includegraphics[height=2cm]{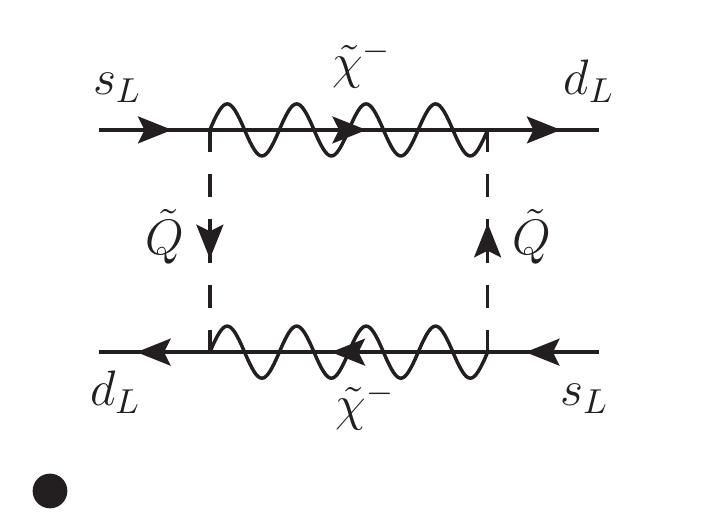}
\caption{Left two diagram: ``Trojan penguin'' box diagrams
  \protect\cite{Kagan:1999iq}. The difference of the two boxes
  contributes to the $\imag A_2$ and increases with the mass difference
  between right-handed up-type ($\tilde U$) and down-type ($\tilde D$)
  squark. $\tilde Q$ denotes a left-handed squark, which is a
  strange-down mixture. Right three diagrams: MSSM contribution to
  $\epsilon_K$. The (second to last) crossed-box gluino diagram cancels
  the middle gluino diagram very efficiently for $m_{\tilde g}\geq 1.5
  m_{\tilde Q}$ \protect\cite{Crivellin:2010ys} and the chargino diagram
  to the right and the boxes with one or two neutralinos
  ($\widetilde\chi{}^0$) become important.\label{fig:trojan}}
\end{figure}

\begin{figure}[t]
\centering
\includegraphics[width=0.75\textwidth]{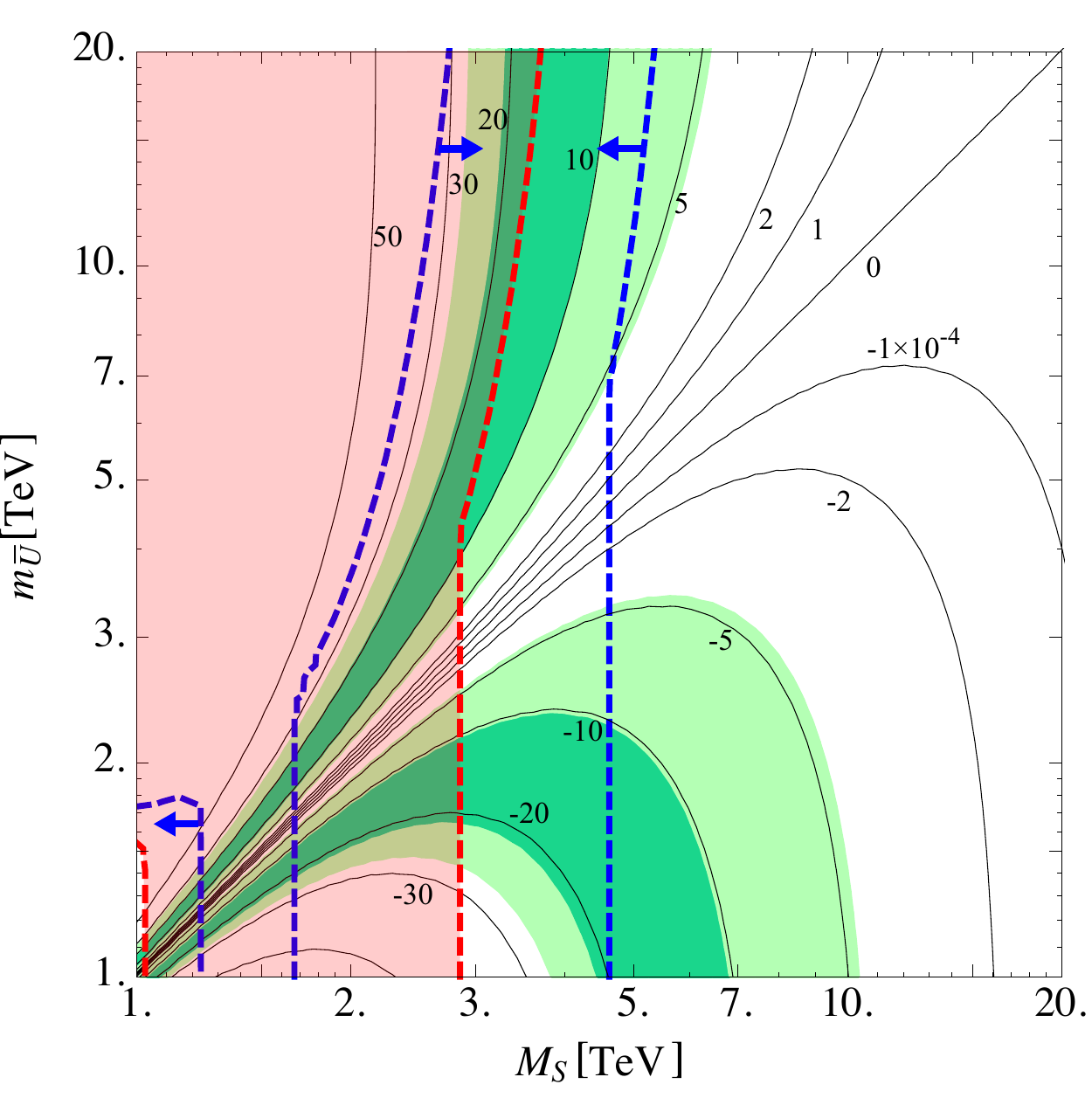}
\caption{Parameter region explaining 
  $\epsilon_K^\prime/\epsilon_K$ while complying with the measured
  $\epsilon_K$ for the point $m_{\tilde g}=1.5M_S$ and $M_S=m_{\tilde
    Q}=m_{\tilde D}$. $M_S$ is a common mass of all superpartners except 
  the gluino and right-handed up squark, labeling the y-axis.
  The lines labeled with negative 
  values of the MSSM contribution 
  $\epsilon_K^{\prime\, SUSY}/\epsilon_K$ correspond to correct
  (positive) solutions if the CP phase $\theta$ is changed from
   $-45^\circ$ to $135^\circ$. 
  The SM prediction
  for $\epsilon_K$ strongly depends on $|V_{cb}|$. The blue (red) 
  lines in both plots delimit the region which complies with 
     $\epsilon_K$ if $|V_{cb}|$ is determined from exclusive (inclusive) 
   $b\to c \ell \nu$ decays. If the exclusive determination is correct, 
   some new physics in $\epsilon_K$ is welcome. In the inclusive case 
   the forbidden region is marked with the red shading. 
   For more details see 
   Ref.~\protect\cite{Kitahara:2016otd}, from which the plots are taken.
   \label{fig:plots}
 }
\end{figure}

The MSSM has the required ingredients to explain $\epsilon_K^\prime$ in
\eq{eq:exp} without conflict with $\epsilon_K$ despite $\delta\gg \tau$.
Moreover, this is possible with squark and gluino masses in the range
3--7\tev, far above the reach of the LHC. The enhancement of
$\epsilon_K^\prime$ is achieved with ``Trojan penguin'' box diagrams
\cite{Kagan:1999iq}, which contribute to $\imag A_2$ through the strong
interaction, as shown in \fig{fig:trojan}. This mechanism involves a
mass splitting among the right-handed up- and down squark and flavour
mixing among the left-handed down and strange squark. The FCNC parameter
is the (1,2) element $\Delta^{LL}_{ds}$ of the left-handed squark mass
matrix, the CP phase is $\theta\equiv\arg(\Delta^{LL}_{sd})$. 
The suppression of $\epsilon_K$ in this MSSM scenario exploits the
mechanism of Ref.~\cite{Crivellin:2010ys}: For 
$m_{\tilde g}\simeq 1.5 m_{\tilde Q}$ the two gluino boxes in 
\fig{fig:trojan} cancel and for $m_{\tilde g} > 1.5 m_{\tilde Q}$
decoupling sets in quickly, so that the MSSM contribution to 
$\epsilon_K$ stays small.
In \fig{fig:plots} the region of sparticle masses capable to explain 
$\epsilon_K^\prime$ is shown for the choice $\theta=-45^\circ$.
This phase maximises the MSSM contribution to $\epsilon_K$ (proportional to
$\sin(2\theta)$), so that it is clear that the suppression of $|\epsilon_K|$
is not caused by a tuning of $\theta$.  

\section{$\mathbf{K\to \pi\nu\bar \nu}$}
The two decay modes $K^+ \to \pi^+ \nu \bar\nu$ and $K_L \to \pi^0         
  \nu \bar\nu$ are remarkable in two respects: On one hand their
  branching ratios can be predicted with high precision, because all
  hadronic effects are under good control. On the other hand the two 
branching ratios are highly sensitive to new physics. 
The SM predictions of the branching ratios are\cite{kpinunu}:
\begin{eqnarray}
\mathcal{B}(K_L \to \pi^0 \nu \overline{\nu})_{\rm SM} &=& 
     (2.9\pm 0.2 \pm 0.0) \times 10^{-11}\,, \nn
 \mathcal{B}(K^+ \to \pi^+ \nu \overline{\nu})_{\rm SM} &=& 
    (8.3\pm 0.3 \pm 0.3) \times 10^{-11}, \label{eq:smk}
\end{eqnarray}
where the first error stems from the CKM elements and the second error
summarises the remaining uncertainties. The experiment NA62 at CERN will
probe $\mathcal{B}(K^+ \to \pi^+ \nu \overline{\nu})$ at the 10\% level
already in 2018 \cite{na62}.  KOTO at J-PARC will, in a first step,
probe $\mathcal{B}(K^+ \to \pi^+ \nu \overline{\nu})$ at the level
around the SM sensitivity \cite{koto1}.  Later KOTO-step2 aims at a
measurement with an error of 10\% as well \cite{koto2}.

In the SM the decay $s\to d \nu\bar \nu$ triggering $K\to\pi\nu\bar\nu$
proceeds through $Z$ penguin and box diagrams similar to those 
constituing $\epsilon_K^\prime$. It is therefore natural to ask, 
whether the new physics which may contribute to $\epsilon_K^\prime$ 
in \eq{eq:exp} will also affect $\mathcal{B}(K \to \pi \nu
\overline{\nu})$ and whether the measurements of these branching
fractions will help to distinguish among different new-physics models. 
Such correlations typically appear in models with $Z^\prime$ bosons
or modified $Z$ couplings \cite{z}. 
\begin{figure}[t]
\centering
\includegraphics[height=3cm]{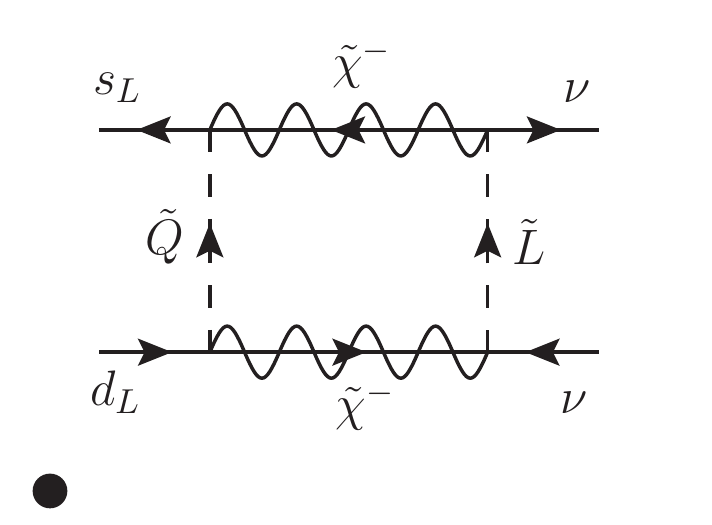}
\caption{Dominant 
  contribution to $K^+ \to \pi^+ \nu \bar\nu$ and $K_L \to \pi^0
  \nu \bar\nu$ in the MSSM scenario of
  Refs.\protect\cite{Kitahara:2016otd,Crivellin:2017gks}. $\widetilde L$
  denotes a charged slepton. Neutralino diagrams sum to a smaller 
  contribution.\label{fig:kpinunu}
}
\end{figure}
In our MSSM scenario there is also a strong correlation between
$\epsilon_K^\prime$ and $\mathcal{B}(K \to \pi \nu
\overline{\nu})$. However, this correlation does not involve
$Z^{(\prime)}$ penguins but instead the box diagrams of
Figs.~\ref{fig:trojan} and \ref{fig:kpinunu}. This diagram is further
correlated to the neutralino (middle and second-to-right) and chargino
(right) diagrams in \fig{fig:trojan}. This correlation now links sizable
enhancements of the $K \to \pi \nu \overline{\nu}$ branching ratios to
excessive effects on $\epsilon_K$, unless one either fine-tunes
$m_{\tilde g}$ to cancel the sum of the two gluino boxes with the
chargino box in \fig{fig:trojan} or tunes the CP phase $\theta$ to
values close to $\pm 90^\circ\;$\footnote{The MSSM contribution to
  $\epsilon_K$ also vanishes for $\theta\approx 0,180^\circ$, but then
  $\epsilon_K^\prime$ cannot be explained.}. If both (fine-)tunings are
restricted to levels below 10\%, one finds
\begin{equation}
\frac{\mathcal{B}(K_L \to \pi^0\nu\overline{\nu})}{\mathcal{B}^{\rm
    SM} (K_L \to \pi^0\nu\overline{\nu}) }\lesssim 1.2\qquad 
\mbox{and}\qquad
\frac{\mathcal{B}(K^+ \to \pi^+\nu\overline{\nu})}{\mathcal{B}^{\rm SM}(K^+
  \to \pi^+\nu\overline{\nu})} \lesssim 1.1, \label{eq:kpred}
\end{equation}
if GUT relations between the gaugino masses are assumed.  Thus the
considered MSSM scenario is very predictive and forbids large effects on
$\mathcal{B}(K \to \pi \nu \overline{\nu})$, although KOTO-step2 may
detect the deviation from the SM prediction. While it is unlikely that
Nature has fine-tuned the gluino mass to minimise the impact on
$\epsilon_K$, the possibility that accidentally $\theta$ is close to
$\pm 90^\circ$ should not be discarded. In this case larger enhancements
than in \eq{eq:kpred} are possible, with the generic strong correlation
between $\mathcal{B}(K^+ \to \pi^+ \nu \overline{\nu})$ and
$\mathcal{B}(K_L \to \pi^0 \nu \overline{\nu})$ found in \cite{blanke},
which holds for any model with flavour mixing only among left-handed
quarks.  (For detailed plots see Ref.~\cite{Crivellin:2017gks}.) Thus
also the scenario with $\theta\approx \pm 90^\circ$ is falsifiable by
combining NA62 and KOTO-step2 data. If, however, the two experiments
both find substantial enhancement following the pattern of
Ref.$\,$\cite{blanke}, the CP phase $\theta$ will be accurately pinned
down to a value near $\pm 90^\circ$.

An interesting prediction of our MSSM scenario is a strict correlation
between $\mathcal{B}(K_L \to \pi^0\nu\overline{\nu})$ and the hierarchy
between the masses $m_{\bar{U}}$,~$m_{\bar{D}}$ of the right-handed
up-squark and down-squark: The (positive) sign of the MSSM contribution
to $\epsilon_K^\prime$ implies
$$\mbox{sgn}\, \lt[ \mathcal{B}(K_L
  \to \pi^0\nu\overline{\nu})-\mathcal{B}^{\rm SM} (K_L \to
  \pi^0\nu\overline{\nu}) \rt] = \mbox{sgn}\, (m_{\bar{U}}-m_{\bar{D}}). $$
Thus a precise measurement of $\mathcal{B}(K_L \to \pi^0\nu\overline{\nu})$
will tell whether the right-handed up squark is heavier or lighter than
the right-handed down squark. 
 
\section*{Acknowledgments}
I thank Andreas Crivellin, Giancarlo D'Ambrosio, Teppei Kitahara, and
Paul Tremper for very enjoyable collaborations and Monika Blanke,
Andrzej Buras, and Amarjit Soni for valuable discussions. I further
thank Teppei Kitahara for carefully proofreading the manuscript.
Support by BMBF under grant no.~05H15VKKB1 is gratefully acknowledged.

\end{document}